\documentclass[prl,showpacs,twocolumn,aps,superscriptaddress,floatfix]{revtex4}
\usepackage{amsmath}
\usepackage{amssymb}
\usepackage{bm}
\usepackage{graphicx}
\usepackage{color}

\DeclareMathOperator{\arcsh}{asinh}

\begin{document}

\title{Metal-insulator transition and phase separation in doped AA-stacked
graphene bilayers}

\author{A.O. Sboychakov}
\affiliation{Advanced Science Institute, RIKEN, Wako-shi, Saitama,
351-0198, Japan}
\affiliation{Institute for Theoretical and Applied Electrodynamics, Russian
Academy of Sciences, 125412 Moscow, Russia}

\author{A.L. Rakhmanov}
\affiliation{Advanced Science Institute, RIKEN, Wako-shi, Saitama,
351-0198, Japan}
\affiliation{Institute for Theoretical and Applied Electrodynamics, Russian
Academy of Sciences, 125412 Moscow, Russia}
\affiliation{Moscow Institute for Physics and Technology (State University), 141700 Moscow Region, Russia}

\author{A.V. Rozhkov}
\affiliation{Advanced Science Institute, RIKEN, Wako-shi, Saitama,
351-0198, Japan}
\affiliation{Institute for Theoretical and Applied Electrodynamics, Russian
Academy of Sciences, 125412 Moscow, Russia}

\author{Franco Nori}
\affiliation{Advanced Science Institute, RIKEN, Wako-shi, Saitama,
351-0198, Japan}
\affiliation{Department of Physics, University of Michigan, Ann
Arbor, MI 48109-1040, USA}

\begin{abstract}
We investigate the doping of AA-stacked graphene bilayers.
Applying a mean field theory at zero temperature we find
that, at half-filling, the bilayer is an antiferromagnetic insulator.
Upon doping, the homogeneous phase becomes unstable with respect to phase
separation. The separated phases are undoped antiferromagnetic
insulator and metal with a non-zero concentration of charge carriers.
At sufficiently high doping, the insulating areas shrink and disappear, and the
system becomes a homogeneous metal. The conductivity changes
drastically upon doping, so the bilayer may be used as a
switch in electronic devices. The effects of finite temperature are also
discussed.
\end{abstract}


%
%
%
%
%
%
%
%
%
%

\maketitle

\textit{Introduction}.--- Controlled metal-insulator (M-I) transitions are
a very useful property for electronic applications of
graphene~\cite{meso_review}. 
Such transitions have been analyzed theoretically
(e.g.,~\cite{MIT})
and experimentally observed in graphene by several groups using different
techniques, e.g., chemical
adsorption~\cite{absor},
thermal
annealing~\cite{therm},
gate-induced M-I
transition~\cite{gate},
and percolation-driven M-I transition in graphene nanoribbons due to
inhomogeneous electron-hole puddle
formation~\cite{puddle}.

Here we study an AA-stacked bilayer of graphene (AA-BLG). The purpose of
this work is to demonstrate that this system, which has been recently
successfully fabricated
\cite{aa_experiment2008,borysiuk_aa2011},
can exhibit a M-I transition upon doping. Further, we will demonstrate that
the required levels of doping are within current experimental capabilities.
Unlike AB-stacked bilayers, the AA-BLG received very modest theoretical
attention
\cite{aa_dft2008,spin-orbit2011,borysiuk_aa2011,aa_adsorbtion2010,
aa_optics_2010,our_aa_prl}.
However, advances in fabrication of AA-stacked bilayers and multilayers
\cite{aa_experiment2008,borysiuk_aa2011}
underscore the need for thorough theoretical investigations.

Tight-binding calculations for AA-BLG
\cite{aa_dft2008,spin-orbit2011}
predict that near the Fermi energy the bilayer has two bands, one
electron-like and one hole-like. These bands have Fermi surfaces, unlike
Fermi points in monolayer graphene and AB-stacked bilayers. An important
feature of the AA-BLG is that the hole and electron Fermi surfaces
coincide. As shown in
Ref.~\onlinecite{our_aa_prl},
if interactions are included, these degenerate Fermi surfaces become
unstable, and the bilayer turns into an antiferromagnetic (AFM) insulator
with a finite gap. This electronic instability is strongest when the bands
cross at the Fermi energy. Impurities or doping shift the Fermi level and
suppress the AFM instability.

Superficially, one may expect that the AFM gap $\Delta$
decreases with doping $x$ and vanishes above some critical value
$x_c$.
However, we will show that the homogeneously-doped state is unstable with
respect to the phase separation into undoped AFM insulator and doped metal.
As the doping grows, the concentration of the AFM insulator shrinks, while
it grows for the metal. Above a certain threshold $x^*$, metallic islands
connect into an infinite cluster, and the percolation-driven
insulator-metal transition occurs, at which point the sample becomes
metallic.

Here we study the electronic properties of the doped AA-BLG in the
framework of the Hubbard-like model used in
Ref.~\onlinecite{our_aa_prl}.
We determine how the gap $\Delta$ depends on $x$ in the
homogeneous state and find the critical value $x_c$, where $\Delta$
vanishes. We further show that at small doping the homogeneous state is
unstable because the compressibility of the system is negative, and find
the doping range where this instability arises. The effects
of non-zero temperature are also discussed.

\textit{The model}.---  The Hamiltonian for $p_z$ electrons of carbon atoms
for the AA-BLG can be written as
\begin{equation}\label{H}
H=H_0+H_{\textrm{int}}-\mu\hat{N},
\end{equation}
where $H_0$ describes electron hopping and
$H_{\textrm{int}}$
is the electron-electron interaction, $\mu$ is the chemical potential, and
$\hat{N}$
is the operator of the total electron number in the system. In the
tight-binging approximation
\begin{eqnarray}\label{H0}
H_0&=&-t\sum_{\langle\mathbf{nm}\rangle i\sigma}
		a^{\dag}_{\mathbf{n}i\sigma}
		b^{\phantom{\dag}}_{\mathbf{m}i\sigma}
\\
\nonumber
&&-
t_0\left(
	\sum_{\mathbf{n}\sigma}
	a^{\dag}_{\mathbf{n}1\sigma}
	a^{\phantom{\dag}}_{\mathbf{n}2\sigma}
	+
	\sum_{\mathbf{m}\sigma}
	b^{\dag}_{\mathbf{m}1\sigma}
	b^{\phantom{\dag}}_{\mathbf{m}2\sigma}
	\right) +
	\textrm{H.c.}
\nonumber
\end{eqnarray}
Here $a^{\dag}_{\mathbf{n}i\sigma}$ and $a^{\phantom{\dag}}_{\mathbf{n}i\sigma}$
($b^{\dag}_{\mathbf{m}i\sigma}$ and $b^{\phantom{\dag}}_{\mathbf{m}i\sigma}$)
are the creation and annihilation operators of an electron with spin $\sigma$
in the layer $i=1,\,2$ on the
sublattice ${\cal A}$ (${\cal B}$) at site $\mathbf{n} \in
{\cal A}$ ($\mathbf{m} \in {\cal B}$).
The amplitude $t$
($t_0$) in Eq.~\eqref{H0} describes the in-plane (inter-plane)
nearest-neighbor hopping. For calculations we will use the values of the
hopping integrals
$t\approx2.57$\,eV, $t_0\approx0.36$\,eV
specific to multilayer AA systems~\cite{Charlier}.
Longer-range hoppings are neglected because these are
small (about or less than 0.1~eV), and we checked that the effects they produce
are negligible (within 1--2\%).

The on-site Coulomb interaction can be written as
\begin{eqnarray}\label{U}
H_{\text{int}}=
\frac{U}{2}\sum_{\mathbf{n}i\sigma}
\left(
	n_{\mathbf{n}i\cal{A}\sigma}
	-
	\frac{1}{2}
\right)
\left(
	n_{\mathbf{n}i\cal{A}\bar{\sigma}}
	-
	\frac{1}{2}
\right)\\\nonumber
+\frac{U}{2}\sum_{\mathbf{m}i\sigma}
\left(
	n_{\mathbf{m}i\cal{B}\sigma}
	-\frac{1}{2}
\right)
\left(
	n_{\mathbf{m}i\cal{B}\bar{\sigma}}
	-\frac{1}{2}
\right),
\end{eqnarray}
where
$n_{\mathbf{n}i\cal{A}\sigma}=
a^{\dag}_{\mathbf{n}i\sigma}a^{\phantom{\dag}}_{\mathbf{n}i\sigma}$,
$n_{\mathbf{m}i\cal{B}\sigma}=b^{\dag}_{\mathbf{m}i\sigma}
b^{\phantom{\dag}}_{\mathbf{m}i\sigma}$, and $\bar{\sigma}=-\sigma$.
It is known that the on-site Coulomb interaction in graphene and other
carbon systems is rather strong, but the estimates available in the
literature vary
considerably~\cite{Ut,U69},
ranging from
$U\sim t$ to $\sim 4t$.
Because of this uncertainty, we will present our results in the form of
$U$-dependent functions, rather than definite estimates.

\begin{figure}
\centering
\includegraphics[width=0.95\columnwidth]{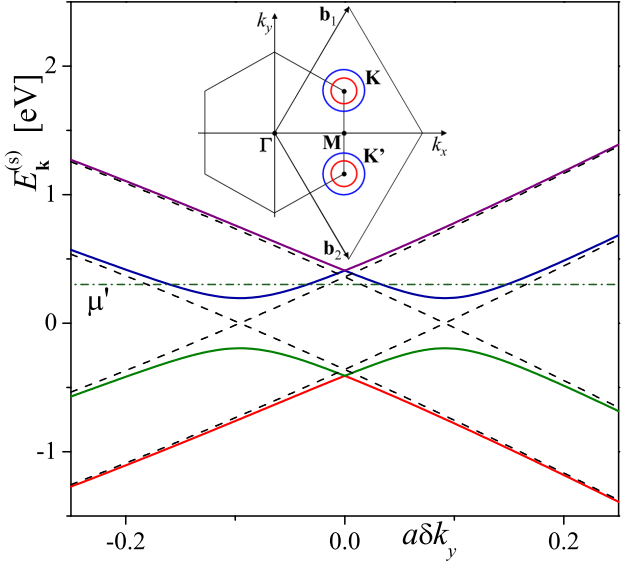}
\caption{(Color online) The band structure for the homogeneous phase
of the AA-stacked bilayer graphene near the ${\bf K}$ point; $\mathbf{k}=\mathbf{K}+\delta k_y\mathbf{e}_y$. The dashed lines show the
non-interacting single-electron bands. At half filling these bands
intersect with each other at the Fermi energy $\mu = 0$.
Adding interactions opens a gap. The mean field bands [see
Eqs.~(\ref{Ek})] are shown by solid lines. With doping, these bands are filled
up to the level $\mu' = \mu - Ux/2$. As a result of doping, the Fermi surface
degeneracy disappears, and we have two Fermi surface components around each Dirac point.
The inset shows the first Brillouin zone (hexagon) and the reciprocal lattice unit cell
(rhombus) of the AA-BLG. Circles around the ${\bf K}$ and ${\bf K}'$ points correspond
to Fermi surfaces of the doped system.
\label{FigSpec0}}
\end{figure}

\textit{Antiferromagnetic state}.---
In the absence of electron-electron coupling,
$U=0$,
and zero doping ($x=0$, which corresponds to half filling) the AA-BLG band
structure is shown in
Fig.~\ref{FigSpec0} by dashed lines.
Two bands pass through the Fermi energy
level near the Dirac points
$\mathbf{K}=2\pi\{\sqrt{3},\,1\}/(3\sqrt{3}a)$
and
$\mathbf{K}'=2\pi\{\sqrt{3},\,-1\}/(3\sqrt{3}a)$, where $a$ is the in-plane carbon--carbon distance.
The chemical potential is
$\mu = 0$,
while the Fermi surfaces are given by the equation
$|f_{\bf k}|=t_0/t$,
where
\begin{equation}\label{f}
f_{\mathbf{k}}
=
1
+
2\exp\!\!\left(
		3ik_xa/2\right)
\cos\!\!\left(k_ya\sqrt{3}/2\right)\,.
\end{equation}
For
$t_0/t \ll 1$,
one can expand the function
$|f_{\bf k}|$
near the Dirac points and demonstrate that the Fermi surface consists of
two circles with radius
$k_{r}=2t_0/(3ta)$
around the Dirac cones
${\bf K}$
and
${\bf K}'$. Upon doping, these Fermi surfaces are transformed into four circles [see the inset in Fig.~\ref{FigSpec0}].
The presence of two bands with identical Fermi surfaces makes the system unstable with respect to spontaneous symmetry-breaking.

Since the unit cell of AA-BLG consists of four atoms, it
is convenient to introduce the bi-spinors
$\psi^{\dag}_{\mathbf{k}\sigma}
=
\left(
	\psi^{\dag}_{\mathbf{k}\cal{A}\sigma},
	\,\psi^{\dag}_{\mathbf{k}\cal{B}\sigma}
\right)$,
with spinor components
$\psi^{\dag}_{\mathbf{k}\cal{A}\sigma}
=
\left(
	a^{\dag}_{\mathbf{k}1\sigma},\,a^{\dag}_{\mathbf{k}2\sigma}
\right)$
and
$\psi^{\dag}_{\mathbf{k}\cal{B}\sigma}
=
e^{-i\varphi_{\mathbf{k}}}
\left(
	b^{\dag}_{\mathbf{k}1\sigma},\,b^{\dag}_{\mathbf{k}2\sigma}
\right)$,
where
$\varphi_{\mathbf{k}}=\arg \{f_{\mathbf{k}}\}$.
The Hamiltonian $H_0$ in this basis is
\begin{equation}\label{Hk}
\hat{H}_{0\mathbf{k}}=-\left(
\begin{matrix}
0&t_0&t|f_{\bf k}|&0\cr
t_0&0&0&t|f_{\bf k}|\cr
t|f_{\bf k}|&0&0&t_0\cr
0&t|f_{\bf k}|&t_0&0\cr
\end{matrix}\right).
\end{equation}

In mean-field, the interaction operator
$H_{\rm int}$,
Eq.~(\ref{U}),
is replaced by a single-particle operator which breaks a certain symmetry
of the system. As it was shown in
Ref.~\onlinecite{our_aa_prl}
the ground state of our model is G-type AFM (that is, the spins on any two
nearest-neighbor sites are antiparallel), for which the spin-up and
spin-down electron densities are redistributed as
$n_{1\cal{A}\uparrow}=n_{2\cal{B}\uparrow}=
n_{2\cal{A}\downarrow}=n_{1\cal{B}\downarrow}=
(1+x+\Delta n)/2$
and
$n_{1\cal{A}\downarrow}=n_{2\cal{B}\downarrow}=
n_{2\cal{A}\uparrow}=n_{1\cal{B}\uparrow}=
(1+x-\Delta n)/2$,
while the total on-site electron density
$n=n_{ia\sigma } + n_{ia\bar\sigma }=1+x$ is the same for any site. The
mean-field interaction Hamiltonian for such phase is
\begin{eqnarray}
\label{HMF}
H^{\textrm{MF}}_{\textrm{int}}\!\!\!
=
\frac{Ux}{2}\hat N
+\Delta
\sum_{\mathbf{k}}
\left(
	\psi^\dag_{\mathbf{k}{\cal A}\downarrow}\hat{\sigma}_{z}\psi_{\mathbf{k}{\cal A}\downarrow}
	-
	\psi^\dag_{\mathbf{k}{\cal A}\uparrow}\hat{\sigma}_{z}\psi_{\mathbf{k}{\cal A}\uparrow}
\right.
\\
\nonumber
\left.
	-
	\psi^\dag_{\mathbf{k}{\cal B}\downarrow}\hat{\sigma}_{z}
	\psi_{\mathbf{k}{\cal B}\downarrow}
	+
	\psi^\dag_{\mathbf{k}{\cal B}\uparrow}\hat{\sigma}_{z}
	\psi_{\mathbf{k}{\cal B}\uparrow}
\right),
\end{eqnarray}
where $\hat{\sigma}_{z}$ is the Pauli matrix, and $\Delta=U\Delta n/2$ is
the AFM gap, which should be found self-consistently.

To find the gap, we solve the corresponding Shr\"{o}dinger equation and
derive the expressions for four electron bands
$E^s(\mathbf{k})$
and eigenvectors
$v_{ia \sigma \mathbf{k} }^{(s)}$
\begin{eqnarray}
\label{Ek}
E^{(1,4)}_{\mathbf{k}}= \mp
\sqrt{\Delta^2+\left(t\zeta_{\mathbf{k}}+t_0\right)^2}\,,
\\
E^{(2,3)}_{\mathbf{k}}= \mp
\sqrt{\Delta^2+\left(t\zeta_{\mathbf{k}}-t_0\right)^2}\,,\nonumber
\end{eqnarray}
where $\zeta_{\mathbf{k}}=|f_{\mathbf{k}}|$. In sublattice ${\cal A}$
for layer $1$, the spin-up wave functions
$\upsilon^{(s)}_{1{\cal A}\uparrow\mathbf{k}}$
are
\begin{equation}
\upsilon^{(s)}_{1{\cal A}\uparrow\mathbf{k}}=\frac12\left[1 -
\Delta/E^{(s)}_{\mathbf{k}}\right]^{1/2}\,.
\end{equation}
The self-consistent equation for the gap is
\begin{equation}\label{EqDelta1}
n_{1{\cal A}\uparrow}=\frac{n}{2}+\frac{\Delta}{U}=%
\sum_{s=1}^{4}\int\!\!\frac{d\mathbf{k}}{V_{\text{BZ}}}\left|\upsilon^{(s)}_{1{\cal A}\uparrow\mathbf{k}}\right|^2%
\!\!\Theta\!\!\left(\mu'-E^{(s)}_{\mathbf{k}}\right),
\end{equation}
where $\mu' = \mu - Ux/2$, $\Theta$ is the Heaviside step-function, and
$V_{\text{BZ}}$
is the volume of the Brillouin zone. The total number of electrons (per
site) $n$ is related to $\mu$ according to
\begin{equation}\label{EqMu1}
n=\frac12\sum_{s=1}^{4}\int\!\!\frac{d\mathbf{k}}{V_{\text{BZ}}}\;
\Theta\!\!\left(\mu'-E^{(s)}_{\mathbf{k}}\right)\,.
\end{equation}

At half-filling, $n=1$, $x=0$, and $\mu'=0$. The lower two bands are filled
while the upper two are empty. Upon electron doping, $x>0$ (hole doping,
$x<0$), $\mu'$ abruptly changes to the new value
$\mu'>\Delta$ ($\mu'<-\Delta$).
Substituting the wave functions
$\upsilon^{(s)}_{1{\cal A}\uparrow\mathbf{k}}$
into
Eq.~\eqref{EqDelta1},
one obtains \begin{eqnarray}
\label{EqDelta2}
1&=&\frac{U}{4t}\int\limits_0^3\!\!d\zeta\,\rho_0(\zeta)\!\!
\left[
	\frac{1-\Theta\left(\displaystyle |\mu'|/t-
	\sqrt{\delta^2+\left(\zeta+\zeta_0\right)^2}\right)}%
{\sqrt{\delta^2+\left(\zeta+\zeta_0\right)^2}}
\right.
+
\nonumber\\
&&\left.
\frac{1-
\Theta\left(
	\displaystyle |\mu'|/t-
	\sqrt{\delta^2+\left(\zeta-\zeta_0\right)^2}
      \right)}
{\sqrt{\delta^2+\left(\zeta-\zeta_0\right)^2}}
\right]\,,
\end{eqnarray}
where $\delta=\Delta/t$, $\zeta_0=t_0/t$, and $\rho_0(\zeta)$ is the
dimensionless density of states $\rho_0(\zeta)=\int
d\mathbf{k}\,\delta(\zeta-\zeta_{\mathbf{k}})/V_{\text{BZ}}$.
Equation~\eqref{EqMu1} implies
\begin{eqnarray}
\label{EqMu2}
|x|&=&\frac12\int\limits_0^3\!\!d\zeta\,\rho_0(\zeta)\!\!%
\left[\Theta\left(\displaystyle |\mu'|/t-\sqrt{\delta^2+\left(\zeta+\zeta_0\right)^2}\right)\right.+\nonumber\\
&&\left.
\Theta\left(\displaystyle |\mu'|/t-\sqrt{\delta^2+\left(\zeta-\zeta_0\right)^2}\right)\right]\,.
\end{eqnarray}

\begin{figure}
\centering
\includegraphics[width=0.95\columnwidth]{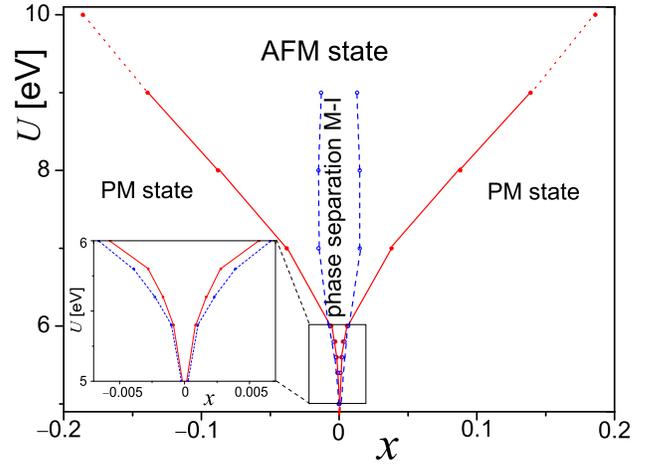}
\caption{(Color online) ($x,\;U$) phase diagram.
Solid (red) lines show the boundary of the uniform AFM state $x_c$.
For large $U$, our mean-field calculations are not quantitatively valid.
To emphasize this, the dotted lines plot $x_c$ for $U>9$~eV.
The dashed (blue) lines show the boundary of
the phase-separated state. The inset shows the magnified phase diagram for
5~eV~$<U<$6~eV.
\label{PhaseDiag}}
\end{figure}

Solving Eqs.~\eqref{EqDelta2} and~\eqref{EqMu2} we obtain $\Delta(x)$
and $\mu(x)$. This can be done analytically if $\Delta_0\ll t,t_0$
($\Delta_0$ is the gap at zero doping). If $\Delta_0$ is small, the value of
$|\mu'|\sim\Delta_0$ is also small, and we can omit $\Theta$-functions in
the first terms in Eqs.~\eqref{EqDelta2} and \eqref{EqMu2}. From these,
we derive
\begin{eqnarray}\label{EqMuAppr}
\nonumber
2\rho_0(\zeta_0)\ln\left(
				\Delta_0/\Delta
			\right)
&\cong&
2\rho_0(\zeta_0)\arcsh\!\left(\delta\zeta/\delta\right)\,,   \\
|x|&\cong&\rho_0(\zeta_0)\delta\zeta\,,
\end{eqnarray}
where $\delta\zeta=\sqrt{(\mu')^2-\Delta^2}/t$.
Solving Eqs.~\eqref{EqMuAppr}, we obtain
\begin{eqnarray} 
\label{DelMu1}
\Delta\!=\!\Delta_0\!\sqrt{1\!-\!|x|/x_c},
\\
\mu\!=\!\Delta_0 \!\left[
		{\rm sgn\,}(x)\!-\!x/2x_c
		\right]\!+\!Ux/2,
\label{DelMu2}
\end{eqnarray} 
where the critical doping $x_c\cong \Delta_0t_0/\pi\sqrt{3}t^2$
(the analytical expression for $\Delta_0$ in the limit $\Delta_0\ll t,t_0$ was
found in Ref.~\onlinecite{our_aa_prl}). We see from
Eq.~(\ref{DelMu1})
that the value of the gap decreases with doping, and
$\Delta=0$, if $|x|\geq x_c$. The curves $\Delta(x)$ are symmetric for electron ($x>0$) and
hole ($x<0$) doping. Next-nearest-neighbor
hopping breaks this symmetry. However, for the parameters
characteristic of graphene systems, the asymmetry of $\Delta(x)$ does
not exceed 1--2\%. The critical doping $x_c$ as function of $U$ is shown
in Fig.~\ref{PhaseDiag}. 
Strictly speaking, 
Eqs.~(\ref{DelMu1},\ref{DelMu2}) 
are not valid for
$\Delta_0\gtrsim t,t_0$.
However, numerical calculations demonstrate that 
Eq.~(\ref{DelMu1}) 
holds true with very high accuracy for any ratio of
$\Delta_0/t$.


\begin{figure}
\centering
\includegraphics[width=0.95\columnwidth]{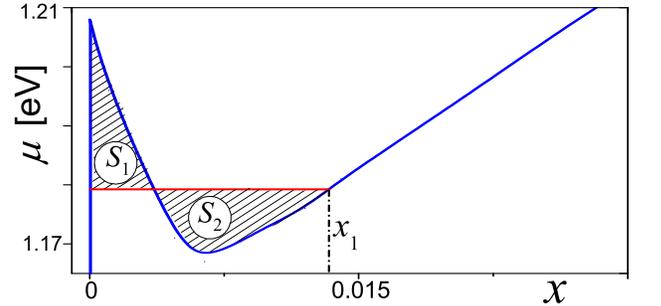}
\caption{(Color online) Chemical potential $\mu$ versus doping $x$ for
the homogeneous state, $U=7$~eV [solid (blue) line]. The horizontal (red)
line shows the Maxwell construction, shaded areas are equal:
$S_1=S_2$.
\label{GapMu}}
\end{figure}

In addition to the usual AFM order parameter, more exotic possibilities are
considered in the literature. For example, doping suppresses the AFM gap,
inducing a canted state~\cite{canting}, in which the angle between
the magnetization vectors in different magnetic sublattices differs from
$180^\circ$. However, our direct numerical calculations of the free energy
show that such canted state is unstable for any doping. Furthermore, the
doped AA-BLG is a typical system with imperfect nesting and, therefore, a
helical AFM state can be induced in
it~\cite{rice}.
This possibility will be analyzed below.

\textit{Phase separation and metal-insulator transition}.---
The chemical potential $\mu$ versus doping obtained from
Eqs.~(\ref{EqDelta2}, \ref{EqMu2}) is shown in Fig.~\ref{GapMu}. Note
that
$\partial\mu/\partial x<0$,
if $|x|$ is small (this result does not depend on the sign of $x$). In
particular, from
Eq.~\eqref{DelMu2}
it follows that
$\partial\mu/\partial x<0$,
if $U/t<\pi\sqrt{3}t/t_0$, which is valid for our choice of parameters. Thus, the compressibility
$\kappa \sim \partial x / \partial \mu$ is negative, indicating the instability of the homogeneous phase
toward phase separation upon doping.
From Fig.~\ref{GapMu}, there are two stable phases with different doping:
$x_0=0$ and $x_1>0$. The value of $x_1$ can be found using the Maxwell
construction~\cite{thermodyn}, according to which the shaded areas in
Fig.~\ref{GapMu} are equal: $S_1=S_2$.
The calculated values of $x_1$ are shown by the (blue) dashed lines in
Fig.~\ref{PhaseDiag} for different $U$s. For the case shown in
Fig.~\ref{GapMu}, $x_1<x_c$, and the uniform system separates into AFM
insulator and AFM metal. For smaller $U$, the situation changes: $x_1>x_c$,
and the co-existing phases are AFM insulator and paramagnetic (PM)
metal (see the inset in Fig.~\ref{PhaseDiag}).

If the doped system were to remain uniform, even small doping would cause
a transition from the insulating magnetic phase to a metallic phase,
magnetic or not. However, the instability of the uniform phase and the
ensuing phase separation delays the transition to the conducting phase
until a finite critical concentration of dopants is reached. Because of this
phase separation, the doped charge segregates into clusters inside the insulating
AFM matrix. The precise structure of such phase depends on a variety of factors:
impurities and defects in the sample or the substrate, the long-range Coulomb
repulsion that arises due to local charge-neutrality breaking~\cite{Cul}, surface tension at the phase
boundaries~\cite{Cul,Surf}, and electron-phonon interactions~\cite{Bian}.
Charge conservation implies that the concentration $p$ of the metallic phase is
$p=|x|/x_1$. The percolative M-I transition occurs if $p$ exceeds some
threshold value $p^*$, which is usually about $0.5$ for 2D systems, and the
corresponding threshold value of doping can be estimated as $|x^*|\sim
0.5\;x_1$.

\textit{Discussion}.--- The most direct and controllable way to switch
AA-BLG from AFM insulator to metal is doping the system with
electron or holes, which could be attained by using appropriate dopants
(e.g., NO$_2$~\cite{absor}, Ca, K~\cite{CaK}), choosing the substrate and applying a gate voltage~\cite{Geim,Kim}
or combining these factors. Our analysis predicts that, for interaction and
hopping parameters values typical for graphene systems, phase separation
exists in the doping range
$0<x<x_1$,
where
$x_1\sim 0.5\!-\!1.5$\%.
Thus, the M-I transition occurs at
$x^*\sim 0.25\!-\!0.75$\%.
For graphene systems, the doping levels $\sim 1\%$  are within the reach of
current experimental techniques such as the adsorption of NO$_2$
gas molecules~\cite{absor,applPL}. Moreover, even higher dopings, necessary
to reach the van Hove singularity, were
achieved~\cite{CaK}. These results suggest that the M-I transition we
discuss in this paper can be realized experimentally.

As mentioned above, we did not include the helical AFM state into our
considerations. Such simplification may be justified. Indeed, the helical
AFM phase is mathematically equivalent to the
Fulde-Ferrel-Larkin-Ovchinnikov (FFLO) state in
superconductors~\cite{fflo,sheehy2007},
which is very sensitive to
disorder~\cite{Takada}
and experimentally difficult to observe. Further, even if the helical state
survives disorder, the phase separation and the M-I transition remain
nonetheless: in such a situation the electrons segregate into insulating
commensurate AFM and metallic helical 
phases~\cite{sheehy2007,rice_phasep_preprint}
with the critical concentration
$x^*$
being slightly different from the values estimated above. At the same time,
the mathematical description 
\cite{rice}
of the helical AFM is fairly involved and cumbersome. Thus, we believe that
at the present stage of this research our simplification of the M-I
transition is warranted.


The above calculations are restricted to the mean field
approximation. To what extent the mean field theory offers a reliable
description of the system? This question was discussed in
Ref.~\onlinecite{kos}
for the usual BCS model and for the BCS-like models with finite spin
polarization in
Refs.~\onlinecite{pilati2008,chevy2006,bulgac2007}.
It is generally agreed that for weak interaction the mean field
calculations are accurate in these situations. In the intermediate-coupling
regime the mean field results remain qualitatively correct. Since the
superconducting systems investigated in these papers are mathematically
equivalent to the AFM, both doped and undoped, we may conclude that our
results are at least qualitatively correct even for moderately high $U$.
Currently, numerical many-body approaches (functional renormalization
group
\cite{frg1,frg2} and
Monte Carlo
\cite{frg1,mc})
demonstrated their usefulness for studies of monolayer and bilayer
graphene. These methods may be used as alternatives to the mean field
approach.

If we want to generalize the formalism for finite $T$, we must remember
that in 2D at $T>0$ no long-range AFM order exists.
However, the short-range AFM order survives up to temperatures
$T^{*}(x)\sim \Delta(x)$.
Indeed, following the approach described in
Refs.~\cite{our_aa_prl,chak,manusakis,schakel}, we obtain the estimate of the
crossover temperature in our model $T^{*}(x)\sim
T_{\text{MF}}(x)\approx0.6\Delta(x)$,
where $T_{\text{MF}}(x)$ is the mean-field transition
temperature~\cite{schakel}. Thus, the crossover temperature is higher than
100~K even if $U$ is as small as 5~eV.

The phase separation can also be destroyed if the temperature exceeds
a certain threshold value $T_{\rm PS}$. To calculate $T_{\rm PS}$ we
have to replace ${1-\Theta\left(\displaystyle\mu'-E_{\bf k}\right)}$
by $f(-E_{\bf k}-\displaystyle\mu')-f(E_{\bf k}-\displaystyle\mu')$
in Eq.~\eqref{EqDelta2} and $\Theta$-functions by the Fermi distributions in
Eq.~\eqref{EqMu2}, [$f(\varepsilon)$ is the Fermi distribution function].
Then, we derive $\mu=\mu(x,T)$ as a function of doping and temperature. If
$T >T_{\rm PS}$, the function $\mu(x,T)$ increases monotonously with $x$.
Our numerical analysis shows that $T_{\rm PS}\gtrsim100$\,K, if $U>5.5$\,eV.

In conclusion, antiferromagnetic order, a metal-insulator
transition, and phase separation are predicted for the doped AA-stacked
graphene bilayer. These effects can be observed at temperatures up to
100~K or even higher.

This work was supported in part by JSPS-RFBR Grant No.~12-02-92100, RFBR
Grant No.~11-02-00708, ARO, Grant-in-Aid for Scientific Research~(S), MEXT
Kakenhi on Quantum Cybernetics, and the JSPS via its FIRST program.  AOS
acknowledges partial support from the Dynasty Foundation and RFBR Grant
No.~12-02-31400.


%

\end{document}